\documentclass{article}

\usepackage[final]{neurips_2019}

\usepackage[utf8]{inputenc} % allow utf-8 input
\usepackage[T1]{fontenc}    % use 8-bit T1 fonts
\usepackage{hyperref}       % hyperlinks
\usepackage{url}            % simple URL typesetting
\usepackage{booktabs}       % professional-quality tables
\usepackage{amsfonts}       % blackboard math symbols
\usepackage{nicefrac}       % compact symbols for 1/2, etc.
\usepackage{microtype}      % microtypography
\title{The tension between openness and prudence in responsible AI research}

\author{
  Jess Whittlestone \\
  Leverhulme Centre for the Future of Intelligence\\
  University of Cambridge \\
  \And
  Aviv Ovadya \\
  The Thoughtful Technology Project \\
}

\begin{document}

\maketitle

\begin{abstract}
This paper explores the tension between openness and prudence in AI research, evident in two core principles of the Montréal Declaration for Responsible AI. While the AI community has strong norms around open sharing of research, concerns about the potential harms arising from misuse of research are growing, prompting some to consider whether the field of AI needs to reconsider publication norms. We discuss how different beliefs and values can lead to differing perspectives on how the AI community should manage this tension, and explore implications for what responsible publication norms in AI research might look like in practice. 

\end{abstract}

\section{Introduction}

Concerns about the societal implications of advances in artificial intelligence (AI) have risen dramatically in recent years. Many organisations and initiatives have also begun to develop principles and declarations to ensure that AI is developed and used responsibly \citep{montreal2018}. One challenge for implementing these principles in practice is that doing so introduces tensions: between different principles, and between principles and other things we value \citep{whittlestone2019}. 

One such tension is that between prudence and openness in AI research. This tension is evident in two core principles of the Montréal Declaration: the prudence principle states that ``when the misuse of an AI systems endangers public health or safety and has a high probability of occurrence, it is prudent to restrict open access and public dissemination'' (p.15) and yet the democratic participation principle states conversely that ``artificial intelligence research should remain open and accessible to all.'' (p.12) This tension raises particularly challenging questions for publication norms in AI research: how should the AI community approach the publication of research which has the potential to be used for harm? \citep{leibowicz2019}

 Our aim in this paper is not to answer the question of when or how AI research should be published, but to facilitate more nuanced discussion of this issue in the AI community by laying out some important considerations. In particular, we suggest the need to better understand both (a) why different groups disagree on this issue, and what beliefs underpin differing perspectives, and (b) what `rethinking publication norms' might actually look like in practice. With a fuller understanding of both of these, we hope the community can begin to explore whether there are ways of rethinking publication norms which carefully balance the tension between openness and prudence. 

\section{The tension between openness and prudence}

The AI research community has historically had strong norms around openness. This is clear in, for example, community backlash against the closed access Nature Machine Intelligence journal \citep{dietterich2018}, and the fact that many top conferences are moving towards open review and demanding code and dataset releases. Openness in research is valuable for many reasons, including ensuring that the benefits of research are distributed widely, and enabling scientific progress by helping researchers build on one another’s work. 

However, concerns about potential harms arising from the misuse of advances in AI research are growing \citep{brundage2018malicious, ovadya2019}. This is not a new problem: many other fields, including biology and computer science more broadly, have wrestled with questions of how to manage research with potential for ‘dual-use’ \citep{atlas2006dual, NDSS2018}. This has prompted some to consider whether the field of AI needs to reconsider norms around how research is published and disseminated.

The AI community currently appears divided on this issue. In February, research lab OpenAI announced that they would not be immediately releasing the full version of their most recent language model, GPT-2, due to “concerns about malicious applications” \citep{radford2019}. While some came out in support of the decision and emphasised the importance of conversation around this issue, many prominent researchers were openly very critical: \citet{lipton2019}, for example suggesting that withholding the model was futile or a disingenuous media strategy to create hype. We do not focus specifically on the case of GPT-2 in this paper, but highlight it here to illustrate how controversial this issue can easily become.

\section{Core disagreements}

Clearly, AI researchers hold varying opinions on how we should balance the tension between openness and prudence in AI research. In order to have a more substantive and productive conversation, it is worth unpacking these disagreements a little more.

In some cases, disagreements may come down to fundamentally different personal values: someone who takes openness in research as a fundamental value or places extremely high priority on it may disagree with any proposal to restrict release on principle. However, there are many different reasons to value both openness and prudence in research, and in most cases we hypothesise that the relative value one assigns to each can be understood in terms of more specific beliefs about why and how each is important. We highlight three types of beliefs which appear particularly central to assessing the relative importance of openness vs. prudence: (a) beliefs about risks; (b) beliefs about efficacy; and (c) beliefs about future needs.

\textbf{Beliefs about risks:} As we have discussed, a central concern related to openness is that research advances could be misused by malicious actors in harmful or even catastrophic ways. But prudence in the release of research also creates its own risks: reduced openness could decrease inclusivity and increase concentration of power, giving disproportionate control over AI advances to a few wealthy research groups and nations. 

\textbf{Beliefs about efficacy}: Different groups may also disagree about how effective restricted release will be at mitigating malicious use in practice. One reason for scepticism is that if there's sufficient motivation to develop a technology, another group is likely to develop and release it anyway. On the other hand, delayed or coordination release might still make a substantial difference if it allows time to build defences against misuse, as is common in computer security \citep{householder2017}.

\textbf{Beliefs about future needs}: If it is likely that more advanced AI systems will be developed soon with much greater potential for harm, then it may be worth beginning to develop forward thinking publication norms and processes, regardless of whether they are needed for current systems. However, if more advanced and potentially harmful capabilities are unlikely for a long time, then developing such processes now may be less valuable.

Focusing on these more specific disagreements, rather than extreme positions of `openness' versus `prudence' may help reduce polarization and enable researchers with differing positions to better understand one another and balance tradeoffs.

%These beliefs all relate to questions that researchers can investigate. We can work to more clearly map out the paths to harm in different areas of AI research to better assess risks from mal-use. We can explore how other fields like biotechnology have managed to avoid the problems power concentration and public confusion (and if they have failed to do so, we can learn something from their failures.) We can also look to analogous cases to get better evidence on the efficacy of different approaches to restricting release (for example, we might explore lessons from responsible disclosure practices in computer security). We can use forecasting and horizon-scanning methods to explore whether near-term advances in capabilities could substantially increase risks of mal-use, and use this to inform decisions made today.

\section{Rethinking publication norms} \label{sec:4}

Different beliefs about risks, efficacy and future needs can lead to differing perspectives on whether the AI community needs to rethink publication norms. But what exactly does it mean to ‘rethink publication norms’?

A number of research groups and labs have already started tackling these questions. One of the most notable public positions in this space comes from Google, which states that ``We generally seek to share Google research to contribute to growing the wider AI ecosystem. However we do not make it available without first reviewing the potential risks for abuse. Although each review is content specific, key factors that we consider in making this judgment include: risk and scale of benefit vs downside, nature and uniqueness, and mitigation options.'' \citep{google2019}

As Google’s emphasis on context-specificity highlights, there are many different options for how research might be published. This section outlines some of these different dimensions and decisions that need to be considered in thinking about publication norms in AI research. Our discussion draws substantially on our understanding of established practices in biology and computer security, which have both dealt with similar issues.

In practice, to think carefully about publication norms the AI community needs to consider (at least) five different questions:

\begin{enumerate}
    \item \textbf{What different options} are there for how research is released?
    \item \textbf{Under what circumstances} should different types of release be used?
    \item \textbf{What processes} should govern how these decisions about release type are made?
    \item \textbf{Who should be involved} in making these decisions?
    \item \textbf{Who or what should manage} (and fund) all of the above?
\end{enumerate}

For example, the question "what different options are there for how research is released?" can be broken down into three different aspects: (1) \textit{content}: what is released, (2) \textit{timing}: when it is released, and (3) \textit{distribution}: where/who it is released to. (See \citet{ovadya2019} for a more detailed breakdown of release options).

%\begin{itemize}
  %  \item \textbf{content (what is released)}: e.g. a fully runnable system; a paper explaining how a system was developed (with varying amounts of detail); a theoretical idea for how a capability might be developed.
    %\item \textbf{timing (when it is released)}: e.g. immediate release; timed release (a specific delay allowing time for mitigation of harms); staged release (systems of increasing levels of power are released successively).
    %\item \textbf{distribution (who it is released to)}: e.g. public access (with varying levels of publicity); `ask for access' where requests are vetted; research groups develop their own trusted communities to share research with informally.
%\end{itemize}

Different combinations of content, timing and distribution options will pose different risks depending on the domain and potential risks of malicious use vs. concentration of power. It is worth considering here different factors that influence the likelihood of a capability being used to cause significant harm: including the awareness and attention of malicious actors; what skills and resources are required to actually deploy a capability in practice; and whether the `return on investment' for using a capability is high enough for adversaries to continue using it in practice.

Even given current publication norms in AI research, the default for publication is often not the most `open' option in each of these categories. Few researchers or labs always publish fully runnable systems based on their research, and they often do not publish the amount of detail - models, data, and source code - that would be required for a straightforward replication. Researchers do not invest equally in media campaigns for every publication, and instead selectively choose which research outputs to distribute to which audiences. Of course, this does not mean that these practices are perfect or should be used to justify reducing openness further. But it does highlight that the research community doess, in practice, already consider different options for how research is released - often for more practical and commercial reasons.

Finally, considering the variety of different options for how research is released highlights the importance of the other questions: who gets to make these decisions, what processes are required, and how all of this can be managed and funded. Both biotechnology and computer security, which have some precedent for restricting release of outputs of potentially harmful research, have established procedures and institutions underpinning these decisions. Biosafety practices include processes for classifying the risk level of different microorganisms, determined by specialist organisations \citep{atlas2006dual}. Similarly, computer security has processes for responsibly disclosing critical information with potential for misuse, which are managed by entities such as Information Security and Analysis Centres \citep{ISACs2018}. Though it is beyond the scope of this paper to discuss in detail what such processes and institutions might look like for AI research, we think these questions are crucially important and would like to encourages future research and experimentation to explore them more thoroughly.

\section{Is it possible to balance prudence and openness?}

Prudence in AI research is not simply a matter of restricting what gets published. Rather, we suggest that being prudent requires establishing norms and processes for assessing the potential risks of research, weighing those risks with potential benefits, and managing any resulting actions chosen to address risks. Though this notion of prudence conflicts with openness in the most absolute sense, it may be possible to establish processes and new supportive institutions which can mitigate the greatest risks of malicious use while retaining many of the most important benefits of openness.

% Prudence in AI research is not simply a matter of restricting what gets published. Rather, we suggest that being prudent requires establishing norms, processes, and potentially institutions for assessing when the risks of research outweigh the benefits, and deciding what to do in those situations. Though this notion of prudence conflicts with openness in the most absolute sense, it may be possible to establish processes which mitigate the greatest risks of malicious use while retaining many of the most important benefits of openness.

For example, the concern that reduced openness may increase power concentration could be substantially mitigated by having well-established processes and institutions governing how publication decisions get made: ensuring that such decisions have legitimacy and that research labs cannot use misuse as an excuse for simply keeping research advances to themselves. Similarly, coordinated approaches to communicating about decisions to restrict or delay release could help prevent public mistrust and confusion.
% TODO: Room to add the example of how instituions can actually help increase opennness and prudence? E.g., a better way to vet researchers and share sensitive models would mean GPT-2 could have been used more braodly earlier.  
In some cases a tradeoff between openness and prudence in research may remain even once we have a better understanding of the risks and benefits of research. In such cases, deeper ethical and philosophical work may be needed, drawing on the work of other fields which have dealt with similar tensions (such as bioethics), to decide what norms and processes should be encouraged.
%Carefully developed distribution mechanisms that allow research to be shared with the relevant research communities could also help ensure that attempts to reduce mal-use do not overly hinder research progress. More generally, better and more widely agreed-upon risk assessment methods for potential mal-use of AI research would help ensure that any decisions to restrict open publication are not taken lightly, and take into consideration a range of information and perspectives. 

Release of outputs is also not the only part of the research process where it may be possible to mitigate the danger of malicious use. For example, perhaps there would be less of a threat to openness if risk assessment could be done at the beginning of the research process to identify areas with potential for harm, leading to decisions not to pursue those directions or to monitor them more closely. This is common in biomedical and behavioural research, where research proposals have to be submitted to institutional review boards (IRBs) which assess whether they pose any risk to human subjects \citep{enfield2008}, though the current implementation of IRBs poses its own significant tradeoffs which need to be thought through. Of course, this is challenging precisely because much of AI research involves developing very general capabilities which can then be used for a variety of different purposes. However, better risk assessment may still be useful for identifying some avenues of research that are not worth prioritising. 

\section{Recommendations}

In order to find a careful balance between openness and prudence in AI research, we suggest that future work should:

\begin{itemize}
    \item \textbf{Aim to better understand risks of misuse} across different areas of AI research, e.g. by conducting threat modelling in collaboration with relevant subject matter experts outside of the AI research community (for example, misinformation security experts for the case of synthetic media research);
    \item \textbf{More thoroughly investigate potential harms of reduced openness} in AI research, including by (a) more substantively engaging with different communities to understand concerns; and (b) better understanding how these harms have arisen, and to what extent they have been dealt with, in other fields;
    \item \textbf{Identify areas where a deeper tension between openness and prudence exists}, even given a better understanding of risks and benefits, and in doing so identify specific questions in need of deeper ethical and philosophical analysis;
    \item \textbf{Explore different options for publication norms and processes} and their real-world impacts in much more detail, using the questions highlighted in \ref{sec:4} as a starting point. This would require publication venues and academic and industry labs to actively experiment with different approaches and share findings on practical challenges, solutions, and outcomes.
\end{itemize}

More generally, we suggest the need to build a community around exploring these issues, with established venues for discussion and learning: for example by running regular workshops specifically on responsible research norms and practices at all major AI conferences.%, with buy-in from leaders in the field. (ADD??)

%\section{Conclusion}

%The Montréal Declaration suggests that both openness and prudence are important for responsibility in AI research: but is it possible to respect both at the same time? We believe that it is, though balancing the two will not be straightforward. Our aim in this paper has been to make the case that this requires developing norms and processes which more carefully balance these two important values. This work needs to begin by more thoroughly understanding the potential risks associated with both openness and prudence in research, and by considering a much wider range of possible approaches to responsible publication norms. This requires considering not just how research gets released but who gets to make these decisions, and what processes and institutions are necessary to legitimise the decision-making process.

%\subsubsection*{Acknowledgments}

\bibliography{references.bib}

\begin{thebibliography}{}

\bibitem[Atlas and Dando, 2006]{atlas2006dual}
Atlas, R.~M. and Dando, M. (2006).
\newblock {The dual-use dilemma for the life sciences: perspectives,
  conundrums, and global solutions}.
\newblock {\em Biosecurity and bioterrorism: biodefense strategy, practice, and
  science}, 4(3):276--286.

\bibitem[Brundage et~al., 2018]{brundage2018malicious}
Brundage, M., Avin, S., Clark, J., Toner, H., Eckersley, P., Garfinkel, B.,
  Dafoe, A., Scharre, P., Zeitzoff, T., Filar, B., et~al. (2018).
\newblock {The malicious use of artificial intelligence: Forecasting,
  prevention, and mitigation}.
\newblock {\em arXiv preprint arXiv:1802.07228}.

\bibitem[Dietterich, 2018]{dietterich2018}
Dietterich, T. (2018).
\newblock {Statement on Nature Machine Intelligence}.
\newblock https://openaccess.engineering.oregonstate.edu/home.

\bibitem[Enfield and Truwit, 2008]{enfield2008}
Enfield, K.~B. and Truwit, J.~D. (2008).
\newblock {The purpose, composition, and function of an institutional review
  board: balancing priorities}.
\newblock {\em Respiratory care}, 53(10):1330--1336.

\bibitem[{European Union Agency for Network and Information Security},
  2018]{ISACs2018}
{European Union Agency for Network and Information Security} (2018).
\newblock {Information Sharing and Analysis Centres (ISACs): Cooperative
  models}.
\newblock Available at
  https://www.enisa.europa.eu/topics/national-cyber-security-strategies/information-sharing.

\bibitem[Google, 2019]{google2019}
Google (2019).
\newblock {Perspectives on Issues in AI Governance}.
\newblock
  https://ai.google/static/documents/perspectives-on-issues-in-ai-governance.pdf.

\bibitem[Householder et~al., 2017]{householder2017}
Householder, A.~D., Wassermann, G., Manion, A., and King, C. (2017).
\newblock {The CERT guide to coordinated vulnerability disclosure}.
\newblock Technical report, Carnegie Mellon University, United States.

\bibitem[Leibowicz et~al., 2019]{leibowicz2019}
Leibowicz, C., Adler, S., and Eckersley, P. (2019).
\newblock {When Is It Appropriate to Publish High-Stakes AI Research?}
\newblock {\em Partnership on {AI}}.

\bibitem[Lipton, 2019]{lipton2019}
Lipton, Z.~C. (2019).
\newblock {OpenAI Trains Language Model, Mass Hysteria Ensues}.
\newblock
  http://approximatelycorrect.com/2019/02/17/openai-trains-language-model-mass-hysteria-ensues/.

\bibitem[{Montréal Declaration on Responsible AI}, 2018]{montreal2018}
{Montréal Declaration on Responsible AI} (2018).
\newblock {Montréal Declaration For A Responsible Development of Artificial
  Intelligence}.
\newblock www.montrealdeclaration-responsibleai.com/the-declaration.

\bibitem[{NDSS}, 2018]{NDSS2018}
{NDSS} (2018).
\newblock {The Network and Distributed System Security 2018 Call for Papers}.
\newblock http://www. ndss-symposium.org/ndss2018/ndss-2018-call-papers/.

\bibitem[Ovadya and Whittlestone, 2019]{ovadya2019}
Ovadya, A. and Whittlestone, J. (2019).
\newblock {Reducing malicious use of synthetic media research: Considerations
  and potential release practices for machine learning}.
\newblock {\em arXiv preprint arXiv:1907.11274}.

\bibitem[Radford et~al., 2019]{radford2019}
Radford, A., Wu, J., Amodei, D., Amodei, D., Clark, J., Brundage, M., and
  Sutskever, I. (2019).
\newblock {Better Language Models and Their Implications}.
\newblock https://openai.com/blog/better-language-models/.

\bibitem[Whittlestone et~al., 2019]{whittlestone2019}
Whittlestone, J., Nyrup, R., Alexandrova, A., Dihal, K., and Cave, S. (2019).
\newblock {Ethical and societal implications of algorithms, data and artificial
  intelligence: a roadmap for research}.
\newblock {\em London: Nuffield Foundation}.

\end{thebibliography}

\end{document}